\documentclass[aps,prl,reprint,groupedaddress,showpacs,amsmath,amssymb,twocolumn,floatfix,preview]{revtex4-1}
\usepackage{graphicx}
\usepackage{bm}
\usepackage{color}
\usepackage[normalem]{ulem}

\newcommand{\dragxy}{$R_{xy}^{drag}$}
\newcommand{\drivexy}{$R_{xy}^{drive}$}
\newcommand{\dragxx}{$R_{xx}^{drag}$}

\newcommand{\CFxy}{$R_{xy}^{CF}$}
\newcommand{\CFxx}{$R_{xx}^{CF}$}
\newcommand{\PFxy}{$R_{xy}^{\parallel}$}

\begin{document}

\title{Excitonic superfluid phase in Double Bilayer Graphene}

\author{J.I.A. Li$^{1}$}
\author{T. Taniguchi$^{2}$}
\author{K. Watanabe$^{2}$}
\author{J. Hone$^{3}$}
\author{C.R. Dean$^{1}$}

\affiliation{$^{1}$Department of Physics, Columbia University, New York, NY, USA}
\affiliation{$^{2}$National Institute for Materials Science, 1-1 Namiki, Tsukuba, Japan}
\affiliation{$^{3}$Department of Mechanical Engineering, Columbia University, New York, NY, USA}

\date{\today}




\maketitle

\textbf{Spatially indirect excitons can be created when an electron and a hole, confined to separate layers of a double quantum well system, bind to form a composite Boson\cite{Loz.75,Pog.77}. Because there is no recombination pathway such excitons are long lived making them accessible to transport studies.  Moreover, the ability to independently tune both the intralayer charge density and interlayer electron-hole separation provides the capability to reach the low-density, strongly interacting regime where a BEC-like phase transition into a superfluid ground state is anticipated\cite{Blatt.62,Loz.75,Loz.76,Pog.77}.  To date, transport signatures of the superfluid condensate phase have been seen only in quantum Hall bilayers composed of double well GaAs heterostructures\cite{Kel.02,Tutuc.04,Wiersma.04,Kel.04}. Here we report observation of the exciton condensate in the quantum Hall effect regime of double layer structures of bilayer graphene.  Correlation between the layers is identified by quantized Hall drag appearing at matched layer densities, and the dissipationless nature of the phase is confirmed in the counterflow geometry~\cite{Kel.04,Su.08}. Independent tuning of the layer densities and interlayer bias reveals a selection rule involving both the orbital and valley quantum number between the symmetry-broken states of bilayer graphene and the condensate phase, while tuning the layer imbalance stabilizes the condensate to temperatures in excess of 4K. Our results establish bilayer graphene quantum wells as an ideal system in which to study the rich phase diagram of strongly interacting Bosonic particles in the solid state.}

In bulk semiconductors, excitons are realized by optically exciting an electron to the conduction band, leaving a hole in the valence band.  Coulomb attraction causes the electron-hole pair to form a bound quasi-particle, referred to as a spatially direct exciton.  While such excitons are easily generated, their spatial proximity leads to recombination on the time scale of a few nanoseconds. By confining the electrons and holes to separate, but closely spaced, 2D quantum wells, strong electron-hole attraction is maintained but recombination is blocked, leading to long-lived excitons. These so-called spatially indirect excitons are predicted to exhibit a rich phase diagram of correlated behaviors, including a type of Bose-Einstein condensation into a superfluid ground state, that should emerge at temperatures much higher than for similar phenomena in atomic gasses\cite{Loz.75,Loz.76,Pog.77}.

\begin{figure*}
\includegraphics[width=1\linewidth]{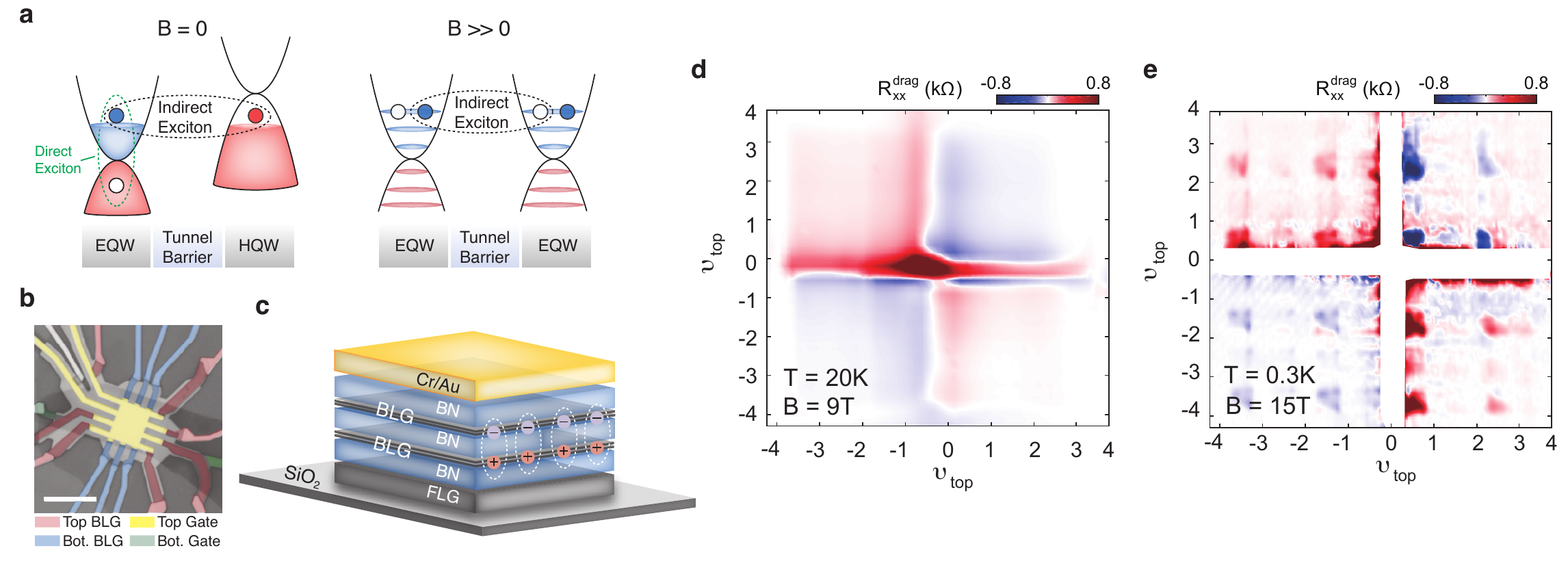}
\caption{\label{fig1}{\bf{Double bilayer graphene}}  (a) Optically excited particle hole pairs combine to form, short-lived, spatially direct excitons.  Electron-hole pairing across a tunnel barrier prevents recombination leading to long lived, spatially indirect excitons. In the QHE regime at large magnetic field, spatially indirect excitons can also result from coupling between partially filled Landau bands. (b) Optical image of a double-bilayer graphene device, with graphite contact and local graphite back gate. The scale bar is $10$ nm. (c) Cartoon cross-section of our device construction (spatially indirect excitons are formed between the two BLG layers). (d) and (e) Longitudinal drag resistance for a device with a tunnel barrier thickness of $d = 5$ nm, as a function of filling factors. (d) Measured at $B = 9$ T and $T = 20$ K. (e) Measured at $B = 15$ T and $T = 0.3$ K. In (e) diverging response near zero density for each layer has been removed for clarity.}
\end{figure*}

At zero magnetic field, transport measurement of the  exciton condensate (EC) phase in coupled electron-hole quantum wells is challenging, owing mostly to the technological challenge of fabricating  matched electron and hole doped 2DEG layers, that are strongly interacting but electrically isolated, while maintaining high mobility~\cite{Gra.91,Sivan.92,Seamons.09,High.12b,Kasprzak.06}.  On the other hand, an equivalent condensate state is possible for identically doped (electron-electron or hole-hole) coupled quantum wells under application of a strong magnetic field.  In the quantum Hall effect (QHE) regime, tuning both layers to half filling of the lowest Landau level can be viewed as populating the lowest band in each layer with an equal number of electrons and holes which then couple across the layers, forming an equivalent system of indirect excitons~\cite{Chak.87,Yosh.89,Eis.04} (Fig. 1a). Indeed with this approach, several measurements have revealed the existence of the EC in GaAs double layers~\cite{Eis.92,Spielman.00,Kel.02,Tutuc.04,Wiersma.04,Kel.04,Nandi.12}, appearing in the QHE regime at total filling, $\nu_{T}=1$ (each layer tuned to $\nu=1/2$) .

For spatially indirect excitons in a magnetic field, $B$, the energy scale of the condensate is conveniently characterized by the effective interlayer separation, $d/\ell_{B}$, where $\ell_{B} =\sqrt{\hbar/eB}$ is the magnetic length, which describes the carrier spacing within a layer, and  $d$ is the thickness of the tunneling barrier separating the layers. Reducing $d$ increases the interlayer coulomb interaction, $e^{2}/\epsilon d$ (and therefore the exciton binding energy), whereas reducing the magnetic length increases the intralayer Coulomb energy, $e^{2}/\epsilon l_{B}$ (increasing interaction energy between the excitons). For GaAs double layers,  in order to prevent interlayer tunneling and maintain sufficiently high mobility, the typical separation between the center of the quantum wells is $d \sim 20$ nm, putting a stringent limit on the achievable effective interlayer separation.  Nonetheless, the EC phase in electron doped GaAs layers is observed to onset for $d/l_{B}\lesssim2$, with a characteristic energy scale of 800~mK ~\cite{Eis.14}.

\begin{figure*}
\includegraphics[width=0.95\linewidth]{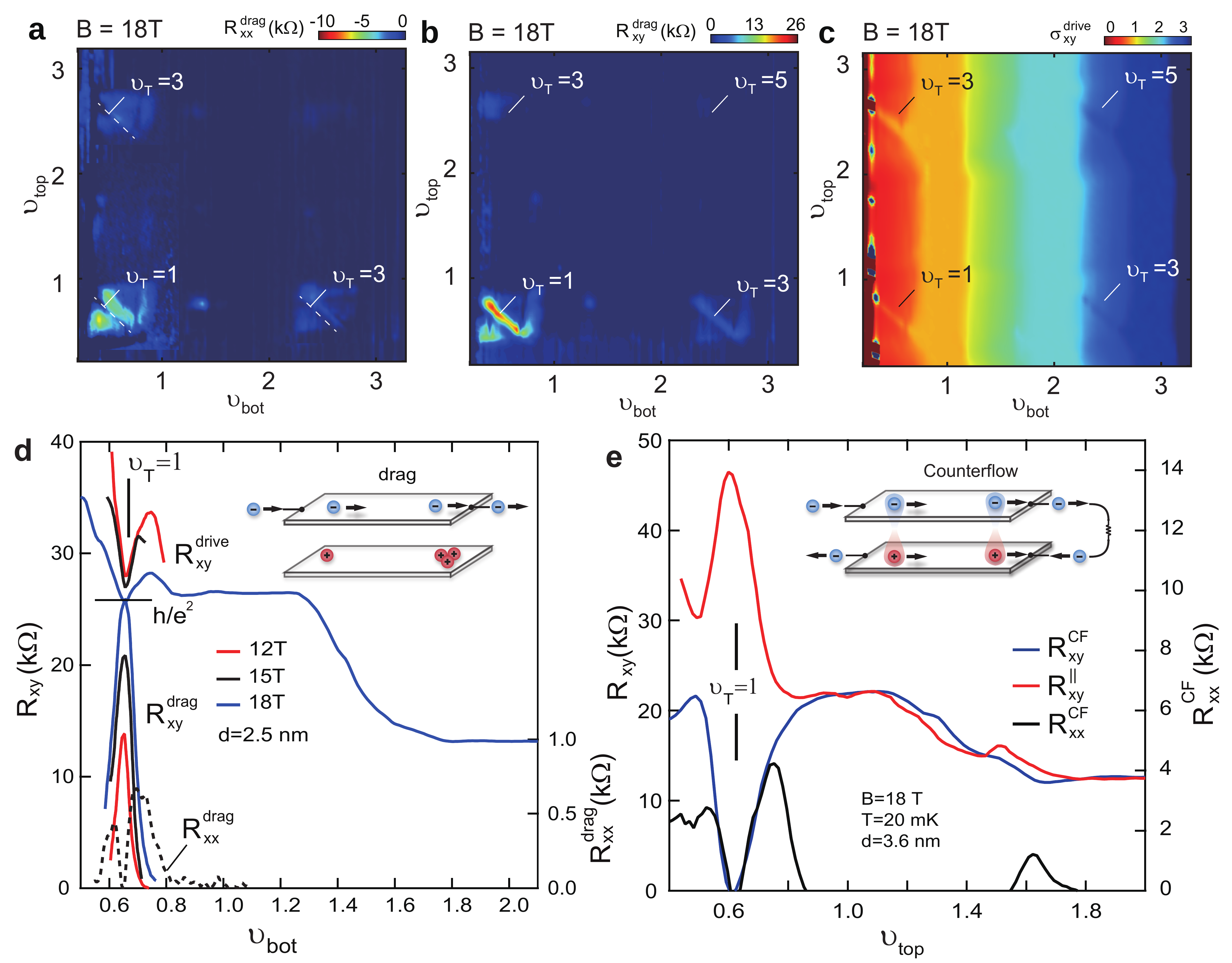}
\caption{\label{fig2} {\bf{Superfluid exciton condensate.}} (a), (b) and (c) Magneto drag (\dragxx), Hall drag (\dragxy) and drive layer Hall resistance (\drivexy)a in the e-e quadrant for device 37 with a tunnel barrier thickness of $d = 3.6$ nm, measured at $B = 18$ T, $T = 20$ mK and $V_{bias} = 0$V. (d) Line cut of \dragxy, \drivexy\ and the magneto-drag $R_{xx}^{drag}$ near $\nu_{T} = 1$ at different magnetic fields for device 45 with a tunnel barrier thickness of $d = 2.5$ nm.  Inset shows the schematic of Coulomb drag measurement. (e) Line cut of counterflow Hall resistance \CFxy\, longitudinal resistance $R_{xx}^{CF}$, and parallel flow Hall resistance \PFxy\ near $\nu_{T} = 1$ measured from the top BLG at $B = 18$ T. Inset shows schematic of the counterflow measurement, which enables transport of charge neutral excitons through the system.}
\end{figure*}

Graphene double layers theoretically promise several advantages over conventional III-V bilayers for realizing the EC phase,  including a carrier density tunable over wide ranges by field effect gating; ambipolar gate response allowing  doping between electrons and holes in each layer; zero layer thickness allowing interlayer spacing down to few nm without significant tunneling ~\cite{Gor.13}, and the possibility of achieving $T_c$ values exceeding cryogenic temperatures\cite{Min.08}. However, while Coulomb drag measurements of double monolayer graphene (MLG) heterostructures have successfully probed the regime of strong interactions (small $d/l_{B}$), no evidence of the EC phase have yet been reported at either zero of finite magnetic field\cite{Gor.13,Kim.11}.  Here we report measurement of double bilayer graphene (BLG) structures in the quantum Hall regime for interlayer separation spanning $2.1$ to $7$ nm, where $d$ is the thickness of the h-BN tunnel barrier. In addition to the potentially more favorable dispersion in comparison with MLG\cite{Hwa.11,Per.13,Zar.14}, the zeroth Landau level (ZLL) of BLG is eight fold degenerate, with the spin and valley isospin degeneracy supplemented by an accidental orbital degeneracy ~\cite{McCann.06}. This multitude of broken symmetry states further expand the phase diagram and enriches the physics of possible superfluid states.

Our devices are assembled using the van der Waals transfer technique ~\cite{Lei.13}. The device geometry includes a local graphite bottom gate, an aligned metal top gate and graphite electrical leads as described in Ref.~\cite{Li.16}. The two BLG are separated by a thin layer of hexagonal Boron Nitride (hBN). Even for the thinnest hBN used ($2.1$ nm) the interlayer tunneling resistance is measured to be larger than $10^9$~ $\Omega$.  Correlation between the layers in the QHE regime is probed by a combination of Coulomb drag ~\cite{Solomon.89}, and magnetoresistance measurements in both counterflow and parallel flow geometries ~\cite{Eis.04,Kel.04,Su.08,Eis.14}. In the drag measurement, current, $I_{drive}$, is sent through the drive BLG layer, while the longitudinal and Hall voltage ($V_{xx}$ and $V_{xy}$) of the drive and drag layers are measured simultaneously. We define the magneto and Hall drag resistance as $R^{drag}_{xx} = V^{drag}_{xx}/I_{drag}$ and $R^{drag}_{xy} = V^{drag}_{xy}/I_{drag}$. Except where indicated, both BLG layers are grounded with no interlayer bias applied across the hBN tunneling barrier. In the counterflow (parallel flow) measurement, equal current is sent through both layers, flowing in the opposite (same) direction, while measuring longitudinal and Hall resistance in each layer ~\cite{Eis.14}  (See SI for schematics of each configuration).

First we examine the Coulomb drag response. At $B=9$~T and $T=20$~K, the longitudinal drag shows conventional behaviour (Fig.~1d), namely a finite response at partial LL filling and dropping to zero when either layer is tuned to a QHE gap.  At the double CNP a large finite response is observed, consistent with previous drag measurements of double MLG and BLG ~\cite{Gor.13,Li.16}. Upon lowering the temperature and increasing the field, we can probe the regime of complete symmetry breaking where the QHE gaps fully developed for all integer filling fractions.  At $B=15$~T and $T=0.3$~K, the overall drag signature diminishes at finite density, but apparently remains robust at certain filling fractions, as shown in Fig.~1e.  Labeling regions of the plot by the coordinates of the bottom and top layer filling fraction, ($\nu_{bottom},\nu_{top}$), an electron-hole asymmetry is apparent. In the electron-electron (e-e) quadrant, magnetodrag is observed whenever there is partial filling of both the $\nu\ = 1$ and $3$ LLs [$(1,1)$,$(1,3)$ $(3,1)$ and $(3,3)$], whereas in the hole-hole (h-h) quadrant it is partially filled $\nu\ = 2$ and $4$ LLs [$(-2,-2)$, $(-2,-4)$, $(-4,-2)$ and $(-4,-4)$]. 

Based on recent understanding of how the 8-fold degeneracy of the zeroth LL in BLG lifts at large magnetic field ~\cite{Yacoby.14,Maher.14,Hunt.16}, we can assign a spin, valley, and orbital index to the symmetry broken states of each layer (see SI).  In doing so, the regions of strong magnetodrag in Fig.~\ref{fig1}e appear to occur only where both layers are in an orbital 0 state, while absent for other combinations.  Magnetodrag due to momentum or energy coupling ~\cite{Hwa.11,Son.12,Schutt.13,Li.16} is expected to vanish in the zero temperature limit, whereas the $0.3$~K response in Fig.~1e exceeds $1$~k$\Omega$ in some regions, suggesting a different origin. One possibility is the formation of indirect excitons between the layers that are not yet phase coherent, resembling the EC precursor reported in GaAs double layers, where $R_{xx}$ first grows to large values with decreasing temperature (or decreasing $d/l_{B}$) before developing a zero valued minimum when the EC fully develops. This interpretation would suggest an orbital selection rule where the EC is stabilized for the zero orbital ground states only. This is consistent with studies in GaAs double layers where the EC has only been observed in the lowest (orbital zero) LL~\cite{Eis.14}.

\begin{figure*}
\includegraphics[width=0.75\linewidth]{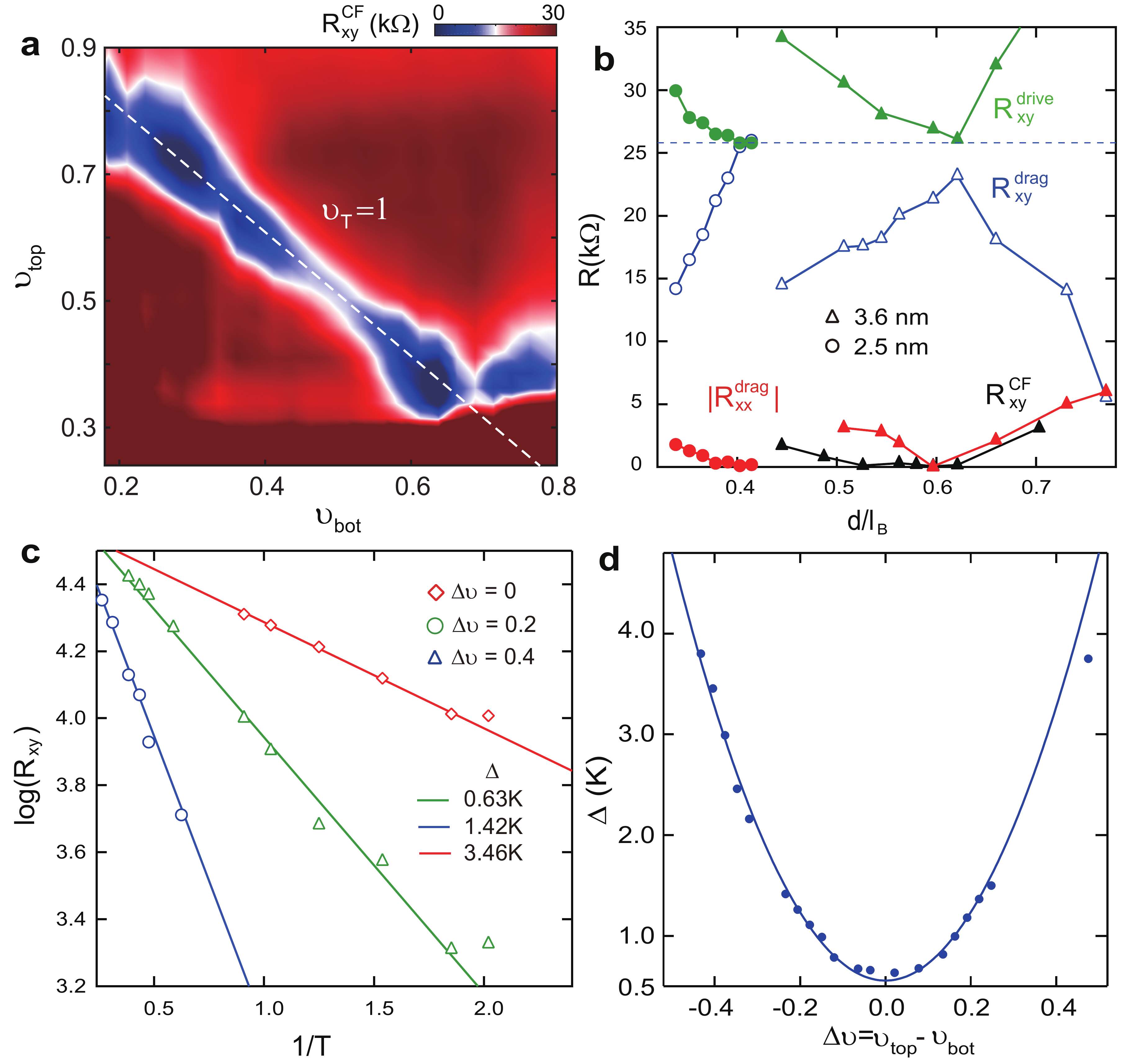}
\caption{\label{fig3} {\bf{Density imbalance.}} (a) \CFxy\ as a function of filling factors for the $\nu_{T} = 1$ state. Points along $\nu_{T} = 1$ can be parametrized by the the interlayer density imbalance $\Delta\nu=\nu_{top}-\nu_{bottom}$. (b)  \dragxy, \drivexy, $|R_{xx}^{drag}|$ and \CFxy\ as a function of effective interlayer separation $d/\ell_{B}$ for device 37 ($d = 3.6$ nm) and 45 ($d = 2.5$ nm). Device 45 shows a much smaller lower critical value of $d/\ell_{B}$ above which full quantization of \dragxy\ and \drivexy\ are observed.  (c)  The temperature dependence of \CFxy\ for device 37, plotted in Arrhenius scale reveals that the energy gap $\Delta$ varies with $\Delta\nu$. (d) The activation gap $\Delta$, (solid circles) appears symmetric with $\Delta\nu$, and fits well to a parabola. \CFxy\ is zero valued in the shaded area, and shows no temperature dependence up to $T \sim 1.2$K. Upper bound critical value of $d/\ell_{B}$ (open circles),  obtained from panel (b) for three different $\Delta\nu$ appears to closely follow the trend of activation gaps}
\end{figure*}

Fig.~2a-c shows the longitudinal magnetodrag (\dragxx), Hall drag (\dragxy), and drive layer Hall resistance (\drivexy),  for a device with interlayer separation of $d=3.6$~nm, measured at $B=18$~T and $T=20$~mK (for simplicity we focus our discussion on the e-e quadrant only, but a complete mapping of the ZLL can be found in the SI). A large response is observed in both \dragxx\ and \dragxy\ following a diagonal line corresponding to total filling fraction $\nu_{T}=1$ ($\nu_{T}=\nu_{top}+\nu_{bottom}$), while at the same time an apparent re-entrance is seen in \drivexy\ at the same total filling. Similar diagonal features are also observed at $\nu_{T} = 3$ and $5$ for the $(1,3)$, $(3,1)$ and $(3,3)$ LLs, corresponding to the same regions identified in Fig. 1e. Fig.~\ref{fig2}d shows the \dragxy\ and \drivexy\ for varying magnetic field measured along a line of varying density in the drive layer. \drivexy\ shows a conventional behaviour with well defined QHE plateaus observed at $\nu_{drive}=1$ and 2, while the magnetodrag is near zero at this sample temperature over most of the density range.  However, when the drive and drag layer densities add to give $\nu_T=1$, \drivexy\ deviates strongly from its single-layer value and instead exhibits a re-entrant behaviour, quantizing to $h/e^2$. At this same total filling, \dragxy\ quantize to this same value. \dragxx\ first rises dramatically in the vicinity of $\nu_{T}=1$ and then goes to a local zero. Quantization of both \drivexy and \dragxy\ at integer total filling, concomitant with a local zero-valued \dragxx,  represents strong collective evidence of the formation of an EC phase ~\cite{Kel.02}.  When interlayer phase coherence is established, electrons are no longer confined to one BLG or the other, but instead exhibit a ``which layer" uncertainty\cite{Eis.14} (this is more easily conceptualized by noting that the EC phase is equivalent to a spontaneous layer-pseudospin ferromagnet forming in the bilayer~\cite{Yang.94}). Therefore even though current is driven through only one layer, it distributes across the double quantum well in the EC state, and the total system manifests features of the $\nu=1$ QHE state, independent of which layer is probed. 


Demonstration that a superfluid EC has truly formed is provided by magnetotransport in the counterflow geometry~\cite{Eis.04}.  In this configuration charge current is carried through the double well system by excitons that are generated (and then annihilated) at the contacts, analogous to the current-carrying cooper pairs in a superconductor ~\cite{Yang.94} (inset Fig. 2e). Being charge-neutral, the excitons feel no Lorentz force even under very large magnetic field,  and a zero Hall resistance is expected ~\cite{Kel.04,Eis.14}.  Indeed, Fig.~\ref{fig2}e shows vanishing counterflow Hall resistance when the same $\nu_{T} = 1$ condition is met. The dissipationless nature of the EC is revealed by the simultaneous zero longitudinal resistance \CFxx. Fig.~\ref{fig2}e also plots Hall resistance from the parallel flow configuration, which is a linear combination of drag and counterflow measurement. The Hall resistance in the parallel flow geometry \PFxy\ shows a prominent peak at $\nu_{T} = 1$,  approaching the quantized value of $2h/e^2$. (This doubling of the quantization is due to the fact that current flows through the double BLG system twice, and \PFxy\ is defined as $V_{xy}/I$ instead of $V_{xy}/2I$.) The stark difference between \CFxy\ and \PFxy\ provides further evidence and confirmation the origin of the $\nu_{T} = 1$ state lies in the strong correlation and interlayer phase coherence between the two BLG layers. 

%

\begin{figure}
\includegraphics[width=1\linewidth]{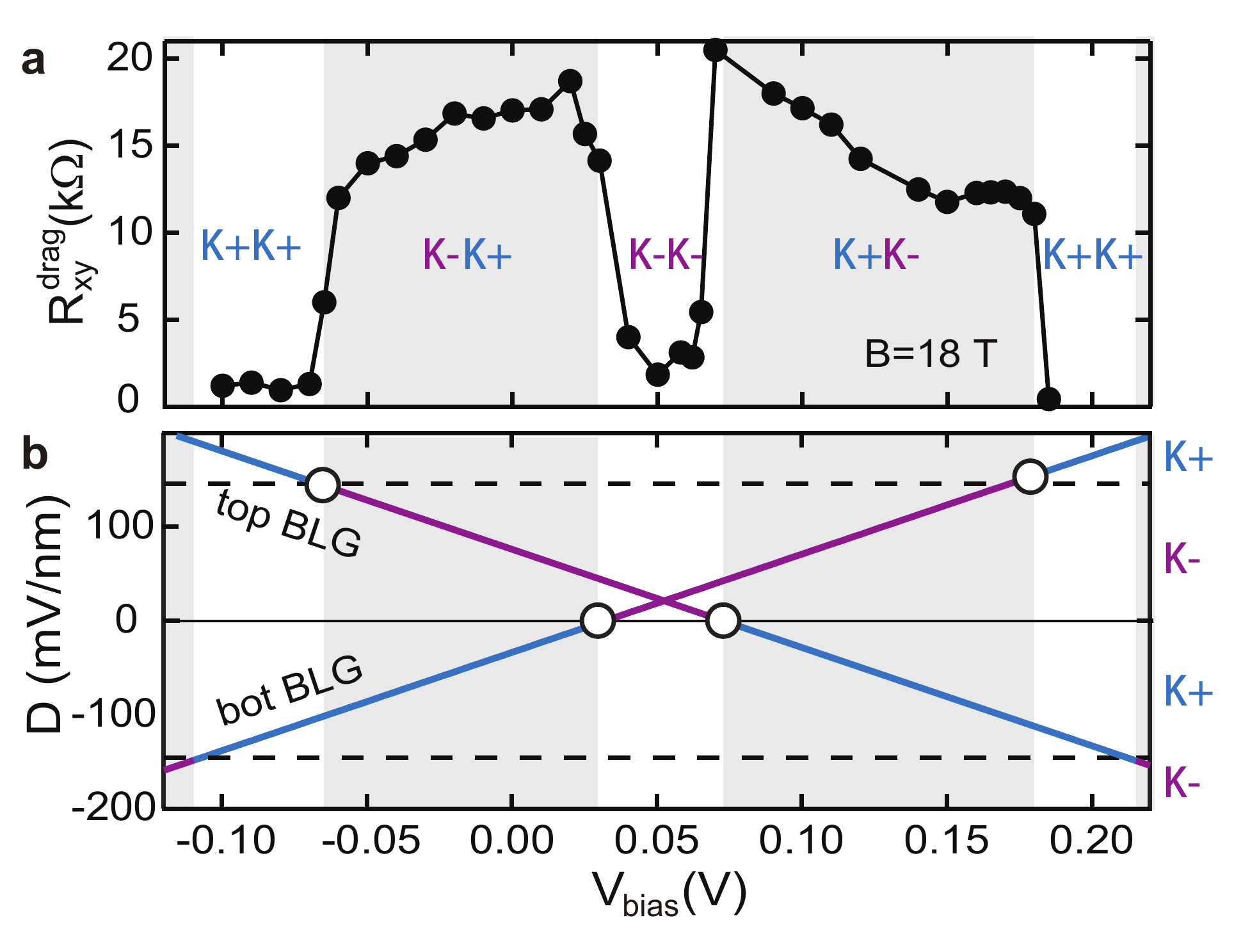}
\caption{\label{fig4} {\bf{Interlayer bias.}} (a) \dragxy\ as a function of interlayer bias, $V_{bias}$. The $\nu_{T}=1$ \dragxy\ peak vanishes and reappears as the EC state displays four different transitions with varying $V_{bias}$. (b) Calculated displacement field $D$ for the two BLG layers as a function of interlayer bias $V_{bias}$ (see SI for detail calculation for the displacement field). The dashed and dotted lines mark critical D-field values for transitions between different valley polarizations, $|K+>$ and $|K->$. The gray shaded area corresponds to opposite valley polarization in the double BLG, $|K+K->$ and $|K-K+>$, whereas the white area indicates the same valley polarization, $|K+K+>$ and $|K-K->$. White circles in lower panel mark transitions in the relative valley order between the layers.    }
\end{figure}


Fig.~\ref{fig3}a shows the counterflow Hall resistance \CFxy\ plotted as a function of filling fractions $\nu_{top}$ and $\nu_{bot}$. The EC state, as evidenced by a zero-valued \CFxy, follows again a diagonal line corresponding to $\nu_{T}=1$.  Along this diagonal the state is described by an interlayer density imbalance, which we parametrize as $\Delta\nu = \nu_{top} - \nu_{bot}$ ($\Delta\nu=0$ only for $\nu_{top}=\nu_{bottom}=1/2$).  To understand the effect of this layer imbalance, we examine the behavior of the $\nu_{T} = 1$ state over a large range of effective interlayer separations and different values of $\Delta\nu$. 

In Fig. 3b we plot the magnitude of \dragxy, \drivexy, $|R^{drag}_{xx}|$ and \CFxy\ versus $d/l_{B}$. For device 37 ($d = 3.6$ nm), nearly quantized \drivexy\ and \dragxy\ together with zero valued \CFxy\, persist only over a narrow range, effectively establishing both an upper and lower critical value for $d/\ell_{B}$.  The upper bound is understood by the requirement to be in the so-called strongly interacting regime (i.e. achieve a minimum effective interlayer interaction), however we note that the critical value $d/\ell_{B}\sim0.6$ is approximately 30$\%$ that reported for GaAs\cite{Kel.04,Nandi.12}. Reducing the interlayer spacing from $3.6$ nm to $2.5$ nm  results in a decrease of the lower critical d/lb (Fig.~3b).  However, we note that this boundary corresponds to approximately the same absolute magnetic field value of approximately $18$ T. This may relate to the minimum magnetic field required to fully lift the ZLL degeneracy (set by sample disorder, which is approximately the same between these two devices).    Alternatively this could be signal of a transition to a new, yet unidentified, phase as $d/\ell_{B}$ tends towards zero.

The minimum value of the \CFxy\  shows activated behaviour with varying temperature, allowing us to deduce an associated gap ~\cite{Kel.04} as a function of the layer imbalance. In Fig. 3d, we plot the activation gap versus $\Delta\nu$.  The data is well fit by a parabolic dependence ~\cite{Jog.02}  with a minimum of $\Delta \sim 0.6$ K near zero density imbalance. This behavior of the energy gap is  consistent with the observation that interlayer density imbalance strengthens the interlayer correlation. This enhancement effect with increasing density imbalance was also observed for the $\nu_T = 1$ phase in GaAs double quantum wells  ~\cite{Champagne.08b}, suggesting the $\nu_T=1$ phase transition in double bilayer graphene could be of the same first order nature as the GaAs double quantum wells ~\cite{Eis.10}. Activated behaviour is observed also for the EC states at $\nu_{T}=3$ and 5, however they exhibit much smaller energy gaps, and are therefore in general less developed compared to the $\nu_{T} = 1$ state. A description of the features observed at these fillings, as well as the equivalent in the e-h quadrant, is provided in the SI. A full analysis of these states however is beyond the present manuscript and will be discussed elsewhere.

Finally, we study the stability of the $\nu_{T} = 1$ state against perpendicular electric field. A voltage bias, $V_{bias}$, is applied to one of the BLG layer (the bottom BLG in this case) to induce the displacement field $D$. The Hall drag signal shows multiple transitions with varying displacement field (Fig. 4a).  The value of the intrerlayer bias at each critical point  shows  good correspondence with D values for which we expect a transition of the valley order in at least one of the bilayers ~\cite{Maher.14,Hunt.16,Yacoby.10}.  Moreover, it appears that the  condensate phase is stabilized (finite drag) when the layers have opposite valley ordering, but suppressed (zero drag) for same ordering (see SI for details). 


In summary, evidence of excitonic superfluidity is observed in double BLG heterostructure in the quantum Hall regime. Quantized Hall drag as well as vanishing counterflow Hall resistance identify the $\nu_{T} = 1$ state to be a counterflow superfluid phase. Measurement over a large range of effective interlayer separation reveals both an upper and lower critical $d/\ell_B$ for the condensate. Studying the $d/\ell_{B}$ dependence as well as the activated energy gap demonstrate that interlayer density imbalance between the two BLG increases the stability of the $\nu_{T} =1$ state, following a $(\Delta\nu)^{2}$ dependence. Within the multicomponent BLG LL spectrum, the condensate phase is observed in a restricted phase space corresponding to orbital quantum number 0 in both layers, coincident with opposite valley polarization.   This work marks the beginning of systematic study of excitonic superfluidity in graphene double layer heterostructures. The capability of engineering and studying the superfluid state in the quantum Hall regime paves the way for realizing such condensate at higher temperature and possibly zero magnetic field.

\begin{acknowledgments}
The authors thank A. Levchenko for helpful discussion.  This work was supported by the National Science Foundation (DMR-1507788). C.R.D acknowledges partial support by the David and Lucille Packard Foundation. A portion of this work was performed at the National High Magnetic Field Laboratory, which is supported by National Science Foundation Cooperative Agreement No. DMR-1157490 and the State of Florida.
\end{acknowledgments}

\end{document}